\documentclass[12pt,a4paper]{article}

\pdfoutput=1  % required by arxiv

\usepackage{xspace}
\usepackage{latexsym}
\usepackage{amssymb}   %

\usepackage{ifthen}
\newboolean{commentsaon}         % to control the top level of comments
\setboolean{commentsaon}{true}
 \setboolean{commentsaon}{false}  % to remove both kinds of comments

\newboolean{commentson} % to control comments
\setboolean{commentson}{true}
\newcommand{\comment}[1]
{\ifthenelse{\boolean{commentson}\AND\boolean{commentsaon}}
   {{\par\noindent\mbox{}{\small\blue[ *** #1 ]\par}\noindent\par}}{}}

\newcommand{\commenta}[1]
{\ifthenelse{\boolean{commentsaon}}
   {{\par\noindent\mbox{}{\small\color[rgb]{0, .5, 0}[ *** #1 ]\par}\noindent\par}}{}}

\usepackage[yyyymmdd]{datetime}

\newcommand{\myhalfmagnification}{0.015}

\usepackage{color}
\newcommand\blue     {\color{blue}}

\usepackage{hyperref}
\usepackage[hyphenbreaks]{breakurl}
\usepackage{pgf}

\newcommand*{\myhfill}{\mbox{}\hfill\mbox{}}

\newcommand*{\HB}{{\ensuremath{\cal{H B}}}\xspace}

\newcommand*{\rr}{\mbox{\bf r}\xspace}
\newcommand*{\cc}{\mbox{\bf c}}

\newcommand*{\jj}{\mbox{\bf j}}
\renewcommand*{\ss}{\mbox{\bf s}\xspace}

\title{The Prolog Debugger and Declarative Programming.  Examples
}

\DeclareRobustCommand{\mytitlecomment}
    {\ifthenelse{\boolean{commentson}\AND\boolean{commentsaon}}
        {\blue\ [with private comments]}{}
    }
\author{%
    W{\l}odzimierz Drabent%
    \\ 
    \normalsize
      Institute of Computer Science,
         Polish Academy of Sciences, \\
    \normalsize
         IDA, Link\"oping University, Sweden
\\
    \normalsize
        {\tt drabent\,{\it at}\/\,ipipan\,{\it dot}\/\,waw\,{\it dot}\/\,pl}
         \\
}
 
\date
     {2020-03-08}
\begin{document}

\maketitle

\begin{abstract}
This paper contains examples for a companion paper 
``The Prolog Debugger and Declarative Programming'', which discusses
(in)adequacy of the Prolog debugger for declarative programming.

  Logic programming is a declarative programming paradigm.  
Programming language Prolog makes logic programming possible, at least to a
substantial extent.
However the Prolog debugger works solely in terms of the
  operational semantics.  So it is incompatible with declarative programming.
The companion paper tries to find methods of using it from the
declarative point of view.  Here we provide examples of applying them.

\end{abstract}

\section{Introduction}

This report contains examples for ``The Prolog Debugger and Declarative
Programming" \cite{drabent.corr.lopstr19}.
The examples present diagnosis of (i.e.\ locating errors in) logic programs
using the Prolog debugger, following the methods (algorithms) described in 
\cite{drabent.corr.lopstr19}.
Erroneous versions of an insertion sort program are used (based on those from
\cite{shapiro.book}).  
For a comparison, we show 
how the same errors are dealt with by declarative diagnosis.
This turns out much simpler than error diagnosis using the Prolog debugger.

Our specification for completeness \cite{drabent.tocl16}
is 
    \[
    \begin{array}{l@{}l}
      {{\it Sp}^0} = 
    S_1  \cup
     \left\{
        {\it insert}( n,l,l' )\in\HB
        \,
        \left|
        \,
        \mbox{\,%
        \begin{tabular}{@{}l@{}}
          $l$ is an ordered list of numbers,
        \\
           $n$ is a number,
        \\
           $l'$ is $l$ with $n$ added and is ordered
        \end{tabular}%
        }
    \right.\right\},
    \end{array}
    \]
where \HB is the Herbrand base (the set of all ground atoms), and
$S_1$ is a specification (for both correctness and completeness) of
$isort$, {\tt>} and {\tt=<}\,,
\[
  \begin{array}{l@{}l}
      {{\it S_1}} =  {} & \,
    \left\{\, {\it i sort}( l,l' )\in\HB
        \,
        \left|
        \mbox{
        \begin{tabular}{l}
         $l'$ is a  sorted permutation
         \\
         of a list $l$ of numbers
        \end{tabular}%
        }
        \right.\right\}
    \ \cup
    \\[2.5ex]
    & \left\{\,
          \begin{array}{@{}c@{}}
            i_1\mathop{\mbox{\tt>}}i_2 \in\HB \\
            j_1\mathop{\mbox{\tt=<}}j_2\in\HB \\
          \end{array}
            \ \left| \   
          \mbox
            {\begin{tabular}{@{}l@{}}
                $i_1,i_2,j_1,j_2$ are integers,\\
                 $i_1>i_2$, $j_1 \leq j_2$
              \end{tabular}%
            }
          \right.\right\}.

  \end{array}
\]
    The specification for correctness is 
    \[
    \begin{array}{l@{}l@{}}
        {\it Sp} = S_1 \cup 
    \left\{\, {\it insert}( n,l,l' )\in\HB
            \,
            \left|
            \mbox{
            \begin{tabular}{@{}l@{}}
              if  $n$ is a number and
            \\ \quad
               $l$ is an ordered list of numbers
            \\
              then $l'$ is $l$ with $n$ added and is ordered
            \end{tabular}%
            }
        \right.\right\}\!.
    \end{array}
\]
So 
${{\it Sp}^0}\subseteq{{\it Sp}}$, and the difference is the set of 
ground atoms ${\it insert}( n,l,l' )$
 in which $n$ is not a number or $l$ is not an
ordered list of numbers.  Such atoms
 may (but do not have to) be answers of a ``right'' program.
We do not bother about the behaviour of $insert$/3 for such arguments.

\section{Diagnosing incorrectness}
Here we deal with this program ({\tt\small inc.isort.pl})
\begin{verbatim}
        isort([X|Xs],Ys) :- isort(Xs,Zs), insert(X,Zs,Ys).
        isort([],[]).

        insert(X,[],[X]).
        insert(X,[Y|Ys],[X,Y|Ys]) :- X =< Y.
        insert(X,[Y|Ys],[Y|Zs]) :- Y > X, insert(X,Ys,Zs).
\end{verbatim}

We use its incorrectness symptom which is an answer  $isort([2,1,3],[2,3,1])$
(for a query $isort([2,1,3],L)$).
For this program examples are shown below of algorithms 
from \cite{drabent.corr.lopstr19} -- Algorithm 2 (finding top-level
search trace),  Algorithms 4, 5 (incorrectness diagnosis) 
from \cite{drabent.corr.lopstr19}, and of declarative diagnosis.

 \pagebreak
\subsection*{Algorithm 2}

\begin{figure}
\hspace*{-2cm}
{%
    \includegraphics
        [viewport = 2cm 12.5cm 19.5cm 26cm, clip]
        {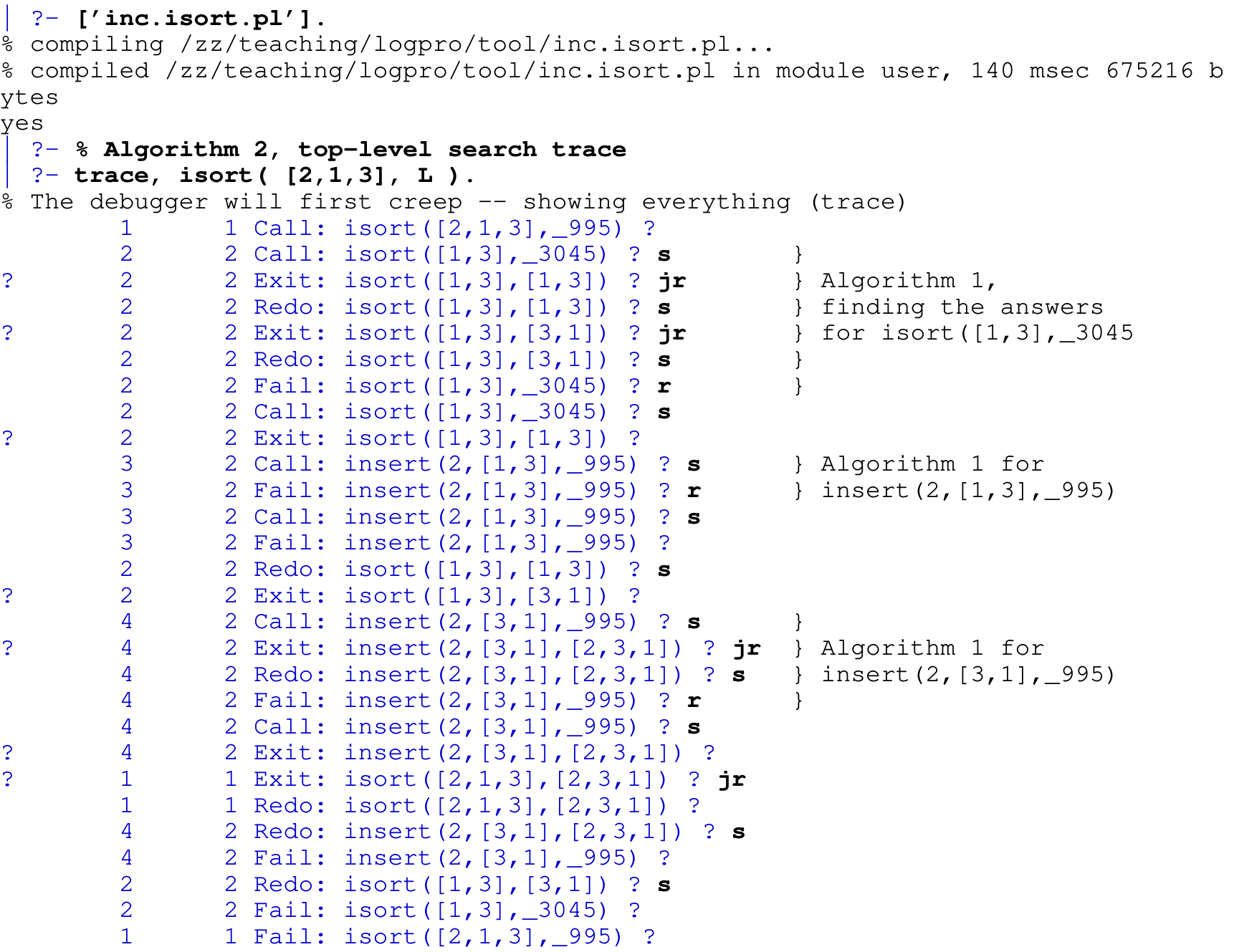}%
}
    \caption{Algorithm 2 (top-level trace) for $isort([2,1,3],L)$}
\label{figAlg2}
\end{figure}

Figure \ref{figAlg2}
 presents constructing a top-level trace for a
query $isort([2,1,3],L)$,  by  Algorithm 2.  Margin comments show the fragments
of the trace obtained by Algorithm 1 (finding all the answers to a query).
The Call and Exit items in these fragments are the top-level trace for the
given initial query.  So the top-level trace is:
\[
\mbox{%
    \tt
    \begin{tabular}{ll}
     \rm query                 &   \rm answers
\\[.5ex]
\hline
\\[-1.5ex]
      isort([1,3],\_3045)     &  isort([1,3],[1,3])
    \\
                             &  isort([1,3],[3,1])
    \\
      insert(2,[1,3],\_995)   & \rm\footnotesize  (none)
    \\
      insert(2,[3,1],\_995)   &  insert(2,[3,1],[2,3,1])  
    \end{tabular}
}
\]

\subsection*{Algorithm 4}

\begin{figure}
\myhfill
{%
    \includegraphics
      [viewport = 2cm 9.6cm 14.8cm 23.33cm,clip]
      {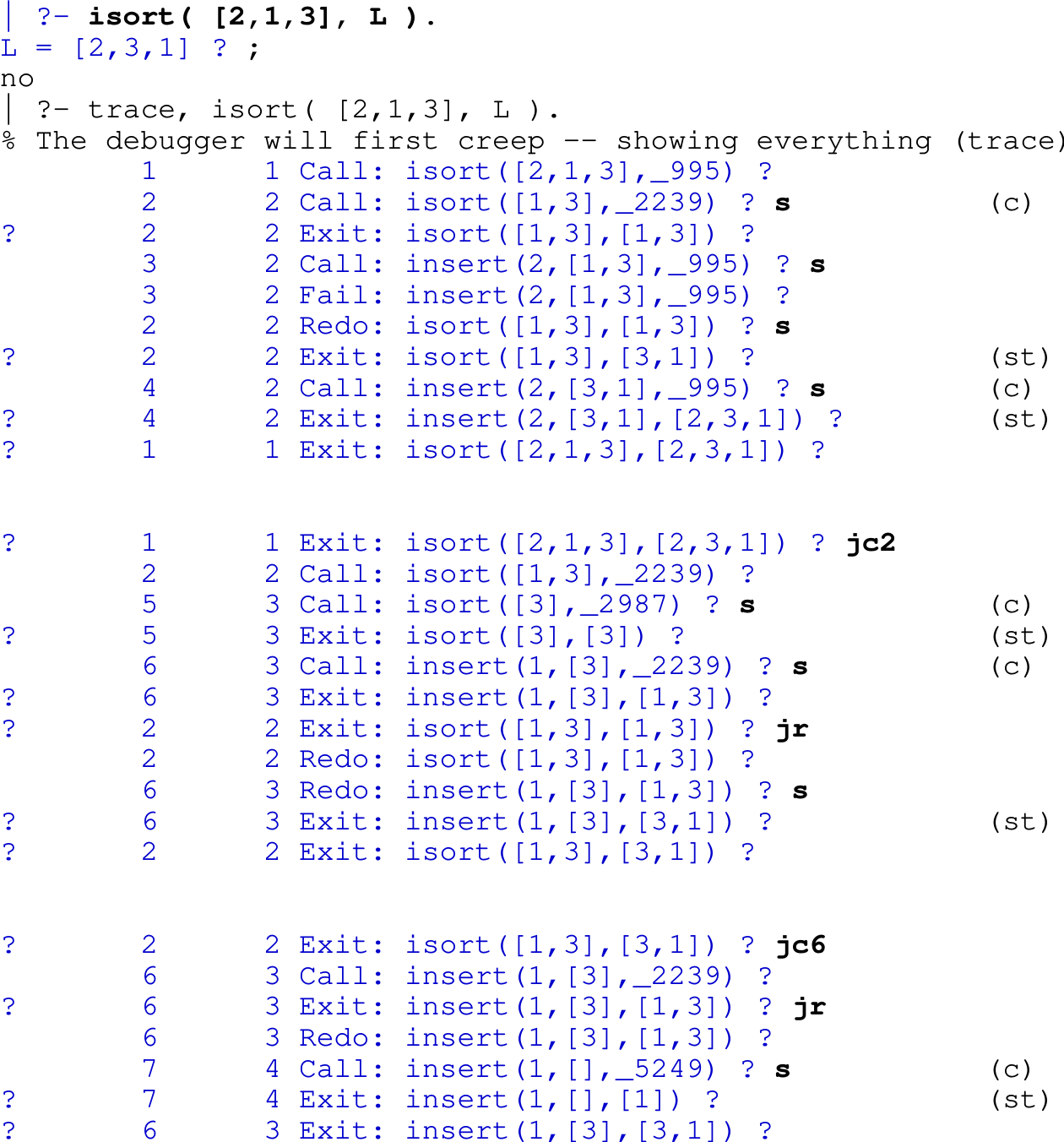}%
}%
\myhfill
\caption{Incorrectness diagnosis according to Algorithm 4}
\label{figAlg4}
\end{figure}

Figure \ref{figAlg4}
 presents diagnosing incorrectness (Algorithm 4)
for query  $isort([2,1,3],L)$.
   The debugging output is split into three parts, each is a construction of
a top-level success trace (Algorithm 3).  The elements of such success trace
are marked with (st) in the margin, the corresponding calls are marked with
(c). 
{\sloppy\par}

First the top-level (search) trace is constructed until arriving to the
incorrect answer  isort([2,1,3],[2,3,1]),  we skip invocations of Algorithm 1. 
Then we pick from the debugger output the top-level success trace for this
atom.

\pagebreak[3]
The success trace consists of two items (formally - of the atoms of items):
\begin{verbatim}
    4      2 Exit: insert(2,[3,1],[2,3,1])
    2      2 Exit: isort([1,3],[3,1])  
\end{verbatim}
The latter is incorrect. 
(Note that, according to the specification for correctness, atom
insert(2,[3,1],[2,3,1])  is correct - we do not bother how items are inserted
into non sorted lists.)

Now command  \jj\cc{\bf2}  leads to the call  $isort([1,3],\_2239)$  that corresponds to
the incorrect atom.  We obtain a relevant fragment of the top-level trace for
this call, and search it for the top-level success trace; its items are
\begin{verbatim}
    6      3 Exit: insert(1,[3],[3,1])
    5      3 Exit: isort([3],[3]) ?  
\end{verbatim}
The former is incorrect, {\bf jc6}  leads to the corresponding call.  The
top-level call trace consists of one item
\begin{verbatim}
    7      4 Exit: insert(1,[],[1])
\end{verbatim}
(the debugger does not show invocations of built-in predicates 
{\tt=<}/2, {\tt>}/2).
As the top-level success trace for the incorrect atom  $insert(1,[3],[3,1])$
consists only of correct atoms, we located an incorrect clause.  It is the
last clause of the program (as  $insert(1,[3],[3,1])$  is an instance of its
head, and - apart from built-ins - it has a single body atom, of which
$insert(1,[\,],[1])$ is an instance).

\newpage
\subsection*{Algorithm 5}

\begin{figure}
\myhfill
{%
    \includegraphics
      [viewport = 2cm 13.3cm 14.8cm 23.33cm,clip]
      {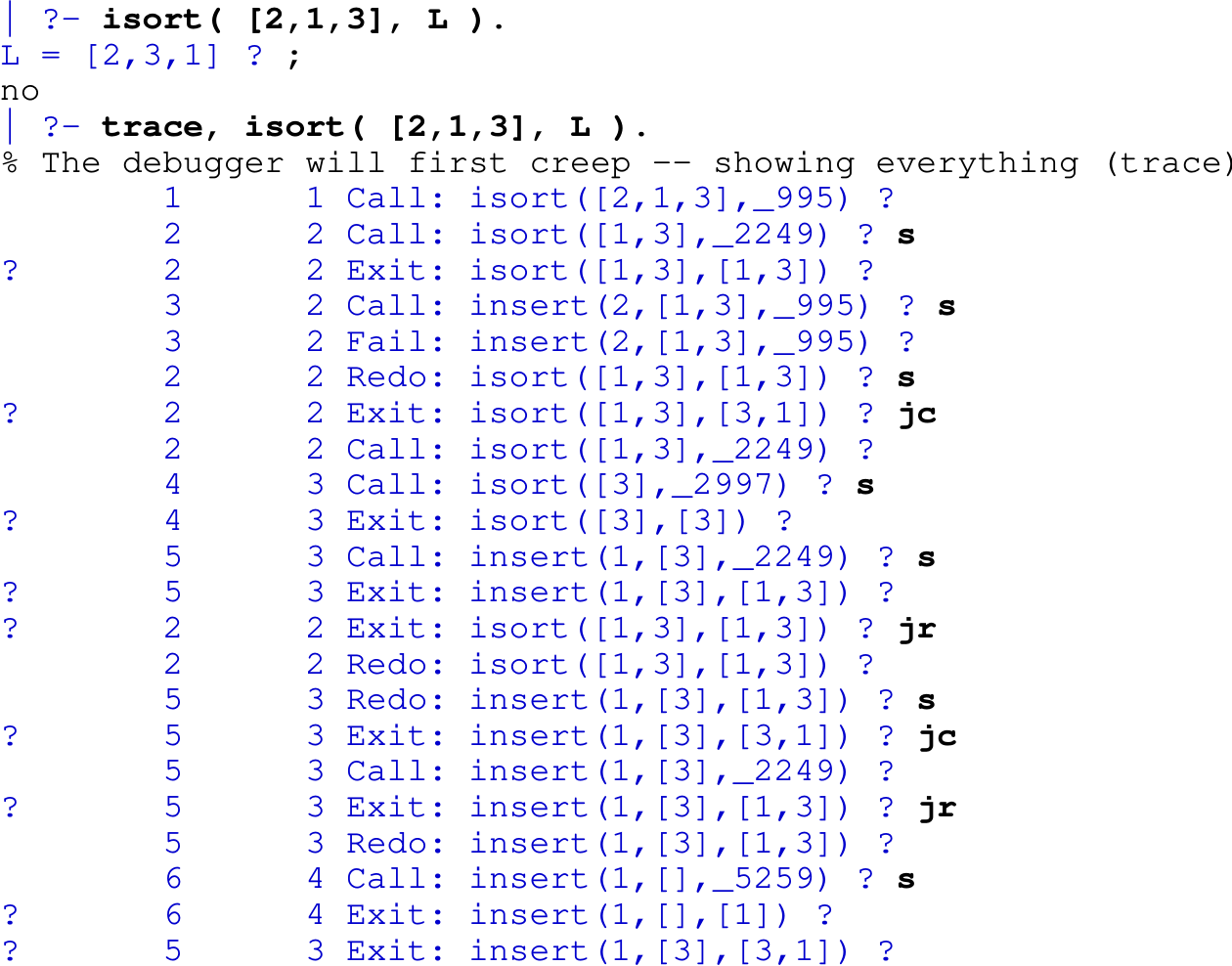}%
}%
\myhfill
    \caption{Incorrectness diagnosis according to Algorithm 5}
\label{figAlg5}
\end{figure}

Figure \ref{figAlg5}  presents applying Algorithm 5 to query
$isort([2,1,3],L)$.   Whenever an incorrect answer is encountered, command
{\bf jc}
is issued to go to the corresponding Call item.  So recursive invocations of
the algorithm were performed for  $isort([1,3],\_2249)$  and  
$insert(1,[3],\_2249)$.
Note that the resulting trace is an abbreviation of that of Algorithm 4.  
The located incorrect clause is the same as in the previous example.

\begin{figure}
\myhfill
{%
    \includegraphics
      [viewport = 2cm 13.3cm 14.8cm 23.33cm,clip]
      {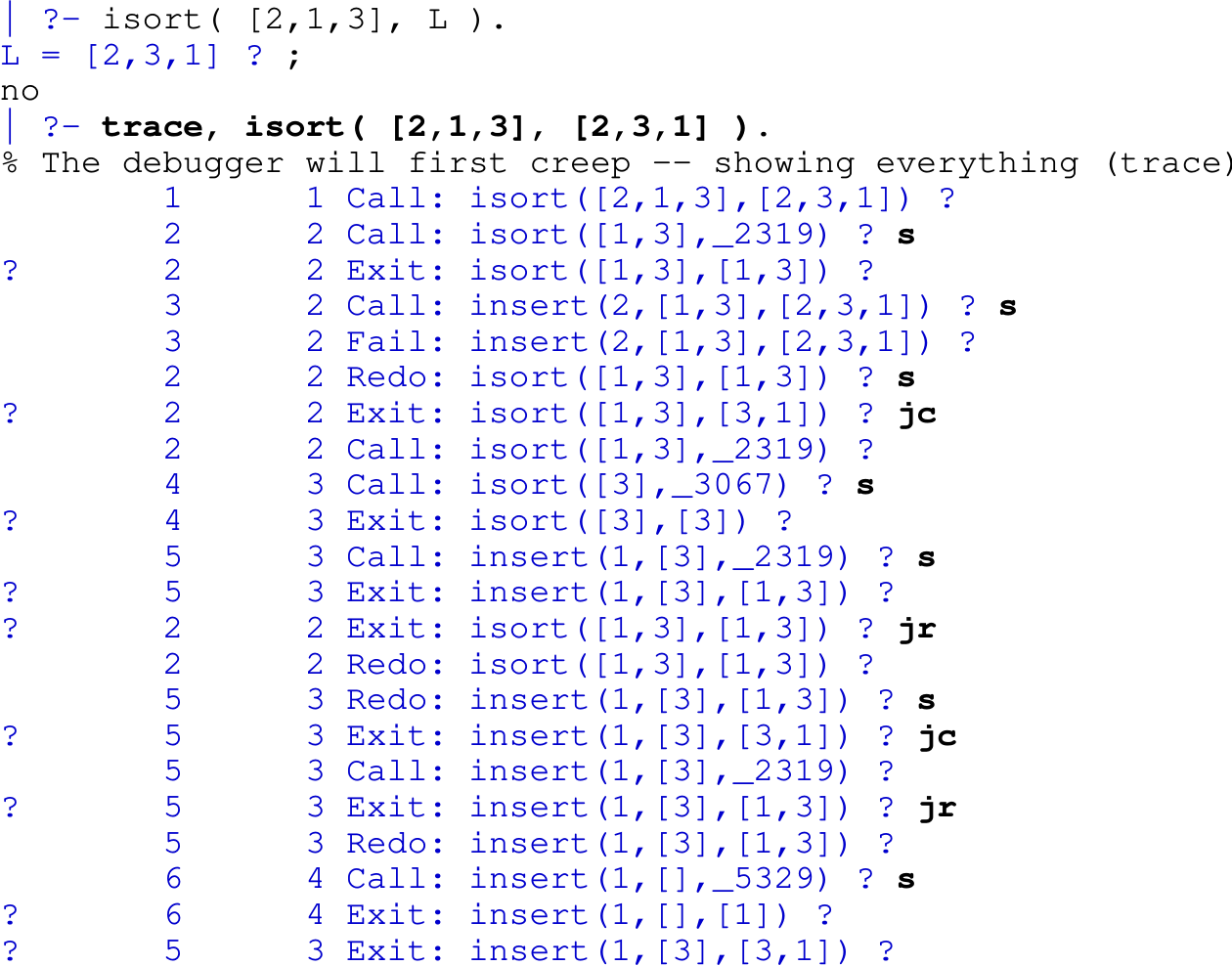}%
}%
\myhfill
    \caption{Incorrectness diagnosis according to Algorithm 5, with the
      incorrect answer as the initial query. }
\label{figAlg5ground}
\end{figure}

Figure \ref{figAlg5ground}  shows that in this particular case the
search of Algorithm 5 is not made shorter by beginning from the incorrect
answer to the considered query (instead of beginning from the query itself).

\begin{figure}
\myhfill
{%
    \includegraphics
        [viewport = 2cm 15.5cm 14.8cm 22.95cm,clip]
      {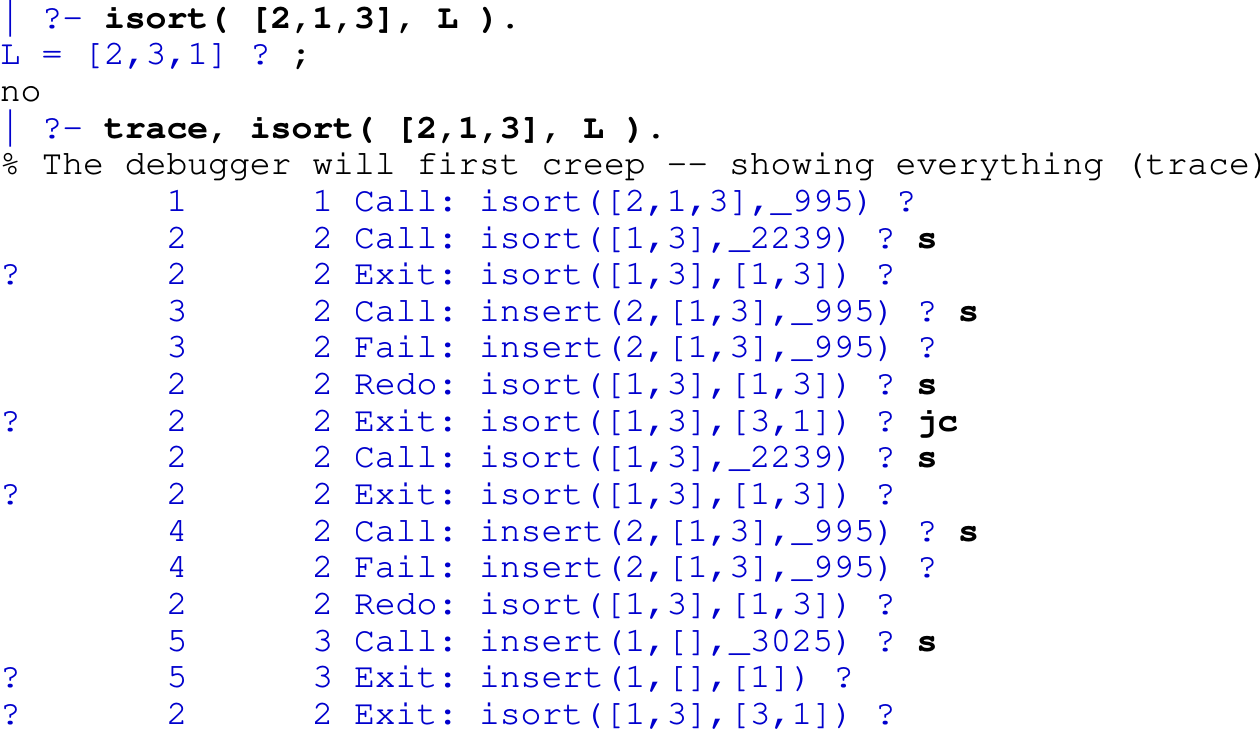}%
}%
\myhfill
    \caption{Incorrect application of Algorithm 5, due to jumping over
      supposedly irrelevant fragments of the top-level trace  }
\label{figAlg5wrong}
\end{figure}

Note that, in the top-level trace for  $isort([1,3],\_2249)$,  we could not jump
over the fragment leading to the first, correct answer.  If we do so, 
the Prolog debugger misses crucial
fragments of the computation leading the the second, wrong answer.
See Figure \ref{figAlg5wrong}.
  After the  \jj\cc\ 
command leading to {\tt Call:\,\,isort([1,3],\_2239)},  command  \ss  was issued, to
immediately obtain the first answer.  As a result,  $insert(1,[3],\_2249)$  was
not traced at all. 
{\sloppy\par}

{\clearpage}

\subsection*
{Declarative diagnosis of incorrectness}

\begin{figure}
\myhfill
{%
    \includegraphics
      [viewport = 2cm 21.5cm 12.7cm 25.7cm,clip]
      {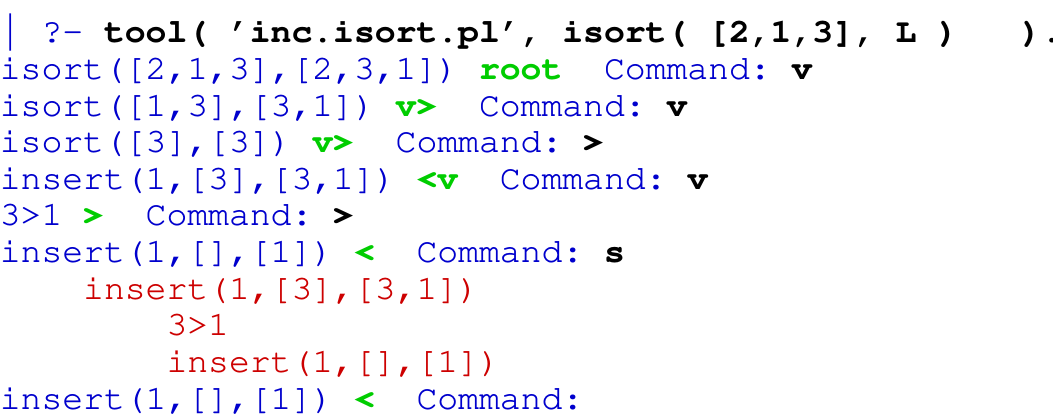}%
}%
\myhfill
    \caption{Declarative incorrectness diagnosis of the same error }
\label{figincDD}
\end{figure}

For a comparison, Figure \ref{figincDD}
presents declarative
incorrectness diagnosis of the program used above, with the same initial
query.  

Note that declarative diagnosis is substantially simpler and more efficient
than that using the Prolog debugger, exemplified above.  In particular, its
log is shorter and much simpler to understand.

Some explanations are due.  The tool used is a prototype browser for proof
trees.  It displays the current tree node (an atom) and asks for a command.
For non-root nodes it displays (in green) the applicable navigation commands.
These commands are:
\[
\begin{tabular}{l@{\ \ --\ \ }l}
   \tt v  &  go to the child,   \\
   \tt <  &  go to the left sibling, \\
   \tt >  &  go to the right sibling,  
\end{tabular}
\]
So we know if the current node is a leaf, a leftmost sibling, or a rightmost
one.  (A command moving to the parent is not displayed and is not used here.)

Declarative
incorrectness diagnosis is performed by navigating the tree looking for an
incorrect child of the last found incorrect node.  Eventually we arrive at
one without incorrect children.  In our case such correct children are $3>1$
and  $insert(1,[\,],[1])$.  This means finding an incorrectness error (an
instance of an incorrect clause).  Command \ss displays (in red) the error. 
(Technically, it displays the parent of the current node with all its
children.)

It is important that the user does not face a strict declarative diagnosis
algorithm, which only asks questions.  Instead, the user decides, what to do.
For instance, when correctness of a given atom is not obvious (say the atom
is big), one may look at its siblings first.  One may explore a subtree, and
then move back; etc.

\section{Diagnosing incompleteness}

Here we use another buggy version of the program, {\tt\small ins.isort.pl}
(in the names of programs ``inc" abbreviates ``incorrectness, and ``ins" 
- ``insufficiency", a synonym for incompleteness):
\begin{verbatim}
        isort([X|Xs],Ys) :- isort(Xs,Zs), insert(X,Zs,Ys).
        isort([],[]).

        insert(X,[Y|Ys],[Y|Zs]) :- X > Y, insert(X,Ys,Zs).
        insert(X,[Y|Ys],[X,Y|Ys]) :- X =< Y.
\end{verbatim}
No answer is obtained for query  $isort([3,2,1], L)$,  thus it is an
incompleteness symptom.

\begin{figure}
\myhfill
{%
    \includegraphics
        [viewport = 2cm 18.15cm 14.8cm 25.6cm,clip]
      {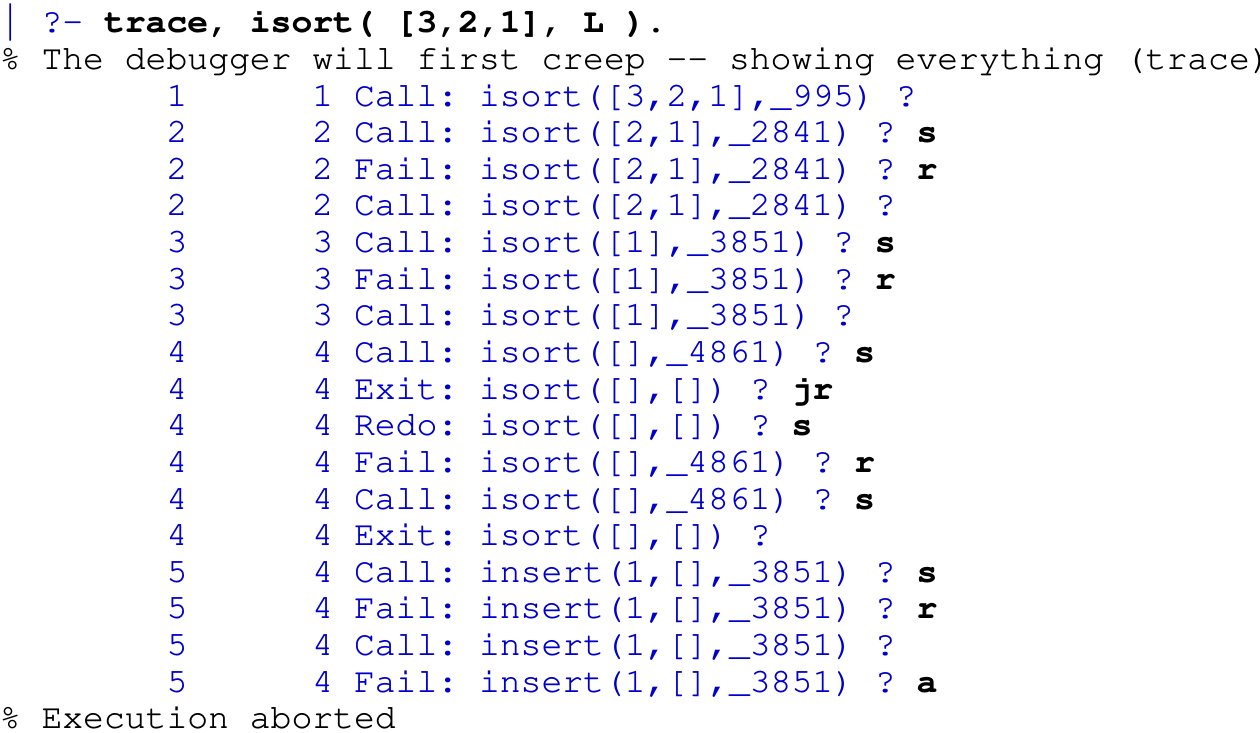}%
}
\myhfill
    \caption{Incompleteness diagnosis according to Algorithm 6.}
\label{figAlg6}
\end{figure}

\subsection*
{Algorithm 6}
Applying algorithm 6 for query
$isort([3,2,1], L)$ is shown in Figure \ref{figAlg6}.
 Construction of a top-level trace (Algorithm 2) is
begun.  The trace contains an incompleteness symptom
\begin{verbatim}
     2      2 Call: isort([2,1],_2841)
\end{verbatim}
(it fails with no answers).  So the construction of the trace is interrupted,
and we begin Algorithm 2 for this query (by commands \rr and ``enter").  Here
\begin{verbatim}
     3      3 Call: isort([1],_3851)
\end{verbatim}
(which fails) is a symptom;  Algorithm 2 is invoked for this query.
The top level trace is:
\[
\begin{tabular}{ll}
  query                &    answers
\\[.5ex]
 \hline
\\[-1.5ex]
 \tt  isort([],\_4861)      &  \tt  isort([],[])
\\
 \tt insert(1,[],\_3851)   &    \rm\footnotesize  (none)
\end{tabular}
\]
and  $insert(1,[\,],\_3851)$  is a symptom, it is also the located
incompleteness error
(as it fails immediately, so its top-level trace is empty, by Algorithm 2).
Procedure insert has been found to be the reason of incompleteness.

\subsection*
{Declarative diagnosis of incompleteness}
\begin{figure}[h]
\myhfill
{%
    \includegraphics
        [viewport = 2cm 23.3cm 10.8cm 25.7cm,clip]
      {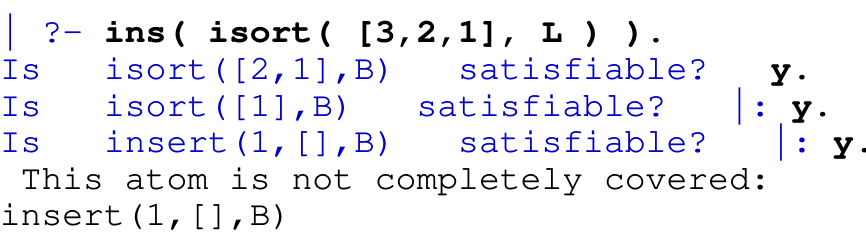}%
}%
\myhfill
    \caption{Declarative incompleteness diagnosis of the same error }
\label{figinsDD}
\end{figure}

For a comparison, we present declarative incompleteness diagnosis for the
same program and symptom. We use a prototype diagnoser of \cite{DNM89}
 (Figure\,\,\ref{figinsDD}).

We should explain the terminology used by the prototype.
An atom $A$ being satisfiable means that ${{\it Sp}^0}\models\exists A$,
where ${{\it Sp}^0}$ is the specification for completeness.
(In other words, this means that $A$ has an instance which is in 
${{\it Sp}^0}$.)
Also, ``not completely covered'' means an incompleteness error
\cite[Def.\,9]{drabent.corr.lopstr19}.

The tool works similarly to Algorithm 6, by constructing top-level search
traces.  Such trace is searched for an incorrectness symptom, by querying the
user about the elements of the trace.
The form of the query is simpler when the trace
element consists of an atom with no answers.  Such simpler queries are asked
first.  As a result, in the presented example there was no need to ask a more
complicated query.%
\footnote{A query about an atom, for which some answers were obtained.
    In our example,
    if the user refuses to answer the query about $insert(1,[\,],B)$ then
    a query is issued about atom $isort([\,],B)$ 
    with a one element set of answers
    $\{ \,isort([\,],[\,])\, \}$.
}

Note the simplicity and efficiency of declarative diagnosis, in comparison 
with that using the Prolog debugger.

\bibliographystyle{alpha}
\bibliography{biblopstr,ja.lopstr19}

\begin{thebibliography}{DNTM89}

\bibitem[DNTM89]{DNM89}
W.~Drabent, S.~Nadjm-Tehrani, and J.~Ma{\l}uszy\'nski.
\newblock {Algorithmic Debugging with Assertions}.
\newblock In H.~Abramson and M.~H. Rogers, editors, {\em Meta-Programming in
  Logic Programming}, pages 501--522. The MIT Press, 1989.

\bibitem[Dra16]{drabent.tocl16}
W.~Drabent.
\newblock Correctness and completeness of logic programs.
\newblock {\em ACM Trans.\ Comput.\ Log.}, 17(3):18:1--18:32, 2016.

\bibitem[Dra19]{drabent.corr.lopstr19}
W.~Drabent.
\newblock The prolog debugger and declarative programming.
\newblock {\em CoRR}, abs/1906.04765, 2019.
\newblock {\small\tt http://arxiv.org/abs/1906.04765}. To appear in {\em
  Logic-Based Program Synthesis and Transformation 29th International
  Symposium, LOPSTR 2019, Porto, Portugal, Revised Selected Papers}, Lecture
  Notes in Computer Science, vol.\,12042. Springer.

\bibitem[Sha83]{shapiro.book}
E.~Shapiro.
\newblock {\em Algorithmic Program Debugging}.
\newblock The MIT Press, 1983.

\end{thebibliography}

\end{document}